\documentclass[preprint, aps, nofootinbib]{revtex4}

\usepackage{graphicx}
\usepackage{dcolumn}
\usepackage{bm}

\def\nslash{\rlap{\hspace{0.02cm}/}{n}}
\def\kslash{\rlap{\hspace{0.02cm}/}{k}}
\def\pslash{\rlap{\hspace{0.02cm}/}{p}}
\def\dslash{\rlap{\hspace{0.08cm}/}{D}}
\def\parslash{\rlap{\hspace{0.02cm}/}{\partial}}
\def\beq{\begin{eqnarray}}
\def\eeq{\end  {eqnarray}}

\def\GeV{{\rm GeV} }
\def\ln{{\rm ln}}
\def\exp{{\rm exp}}
\def\non{\nonumber}

\def\as{\alpha_s}

\def\lqcd{\Lambda_{\rm QCD}}

\newcommand\npb{Nucl.\ Phys.\ B }

\newcommand\plb{Phys.\ Lett.\ B }

\begin{document}

\title{ Sudakov form factor in effective field theory }

\author{Zheng-Tao Wei}
\email{zwei@uv.es} %
\affiliation{ Departamento de F\'{\i}sica Te\'orica, Universidad
  de Valencia, \\ E-46100, Burjassot, Valencia, Spain }

\begin{abstract}
We discuss the Sudakov form factor in the framework of the
soft-collinear effective theory. The running of the short distance
coefficient function from high to low scale gives the summation of
Sudakov logarithms to all orders. Our discussions concentrate on the
factorization and derivation of the renormalization group equation
from the effective theory point of view. The intuitive
interpretation of the renormalization group method is discussed. We
compared our method with other resummation approaches in the
literatures.

\end{abstract}


\maketitle

\newpage
\section{Introduction}

Most high energy processes contain several energy scales which
complicate the analysis in perturbation theory. One classic example
is the elastic form factor of an elemental particle (such as quark
or electron) at large momentum transfer $Q$ \cite{Sudakov}. This
asymptotic form factor is usually called Sudakov form factor. In
one-loop corrections to the Sudakov form factor, a double-logarithms
like $-g^2{\rm ln}^2 \frac{Q^2}{m^2}$ with $m$ a low mass scale will
appear. For the case $Q\gg m$, the large double-logarithms spoil
convergence of the perturbative expansion even if the coupling
constant $g$ is small and they should be resummed to obtain a
well-behaved expansion in perturbation theory. A lot of theoretical
attempts had been made to sum the series $\as^n{\rm ln}^m Q^2(m\le
2n)$ to all orders \cite{Sudakov, resummation, CPQ, Collins,
Korchemsky, Korchemsky2}. All the methods found that the Sudakov
form factor exponentiates and damps rapidly when $Q$ approaches
infinity.

We refer to the summation of double logarithms as Sudakov
resummation. Sudakov form factor is very interesting in theory
because it provides the simplest example to explain basic ideas of
the Sudakov resummation. The earliest treatment in \cite{Sudakov}
introduces the leading double-logarithmic approximation method which
chooses the most important contributions of Feynman diagrams and
then sum them to all orders. Most other methods utilize the standard
renormalization group (RG) technics, such as in \cite{Collins,
Korchemsky}. The central ingredients in them are factorization and
renormalization group equation, although the detailed technics
involve different emphasizes. The separation of the form factor into
the hard, collinear and soft parts for each momentum region leads to
evolution equations and consequently to Sudakov resummation. In
\cite{CLS}, the authors point out a close relation between the
factorization and the matching process in effective field theory.

The effective field theory provides a simple and powerful method to
study processes with several disparate energy scales. Recent
interests on Sudakov resummation come from the study by using a
soft-collinear effective field theory (SCET). SCET is a theory
proposed for collinear and soft particles to simplify the analysis
for the processes with highly energetic hadrons \cite{B1, B2}. This
SCET is a development of the early large energy effective theory
\cite{DG} which includes the collinear quark and soft gluons only.
It was shown in $B\to X_s \gamma$ decays that the summation of the
Sudakov logarithms is much simpler in the effective field theory
than the analysis in the full theory \cite{B1}. This method of
summing double-logarithms had been extensively used in exclusive B
decays such as in \cite{BHLN, SS}.

The baic idea for summing the large logarithms in the effective
field theory can be expressed through an example of B meson decay
\cite{Buras}. One-loop corrections are enhanced by large logarithm
$\ln(m_W/m_b)$ with $m_W, m_b$ the mass of W-boson and b-quark. We
separate the large logarithm $\ln(m_W/m_b)$ into the hard $\ln
(m_W/\mu)$ and the soft $\ln(\mu/m_b)$ parts. The effective field
theory integrates out the heavy $W$-boson and the hard gluons with
virtualities between $m_W$ and a renormalization scale $\mu$. This
process gives an
effective Hamiltonian %
\beq %
H_{\rm eff}=k~C(\mu)~{\cal O}(\mu),
\eeq %
where $k$ is an aggregate for the weak coupling and the CKM matrix
elements. For illustration, we consider the case of single operator.
The $\mu$-dependence of $C(\mu)$ cancels $\mu$-dependence of the
hadronic matrix element of four-fermion current operator
$\langle{\cal O}\rangle (\mu)$. The freedom of
choosing $\mu$ gives evolution equations %
\beq \label{eq:RGE}%
\mu\frac{dC(\mu)}{d\mu}={\gamma(g)}C(\mu), ~~~~~~~ %
\mu\frac{d\langle{\cal O}\rangle(\mu)}{d\mu}=
  -{\gamma(g)}\langle{\cal O}\rangle(\mu).
\eeq %
The anomalous dimension $\gamma(g)=\frac{1}{Z_{\cal
O}}\frac{dZ_{\cal O}}{d\ln \mu}$ is determined by the
renormalization property %
\beq %
C=Z_{\cal O}C^{0},~~~~~~~~ {\cal O}^{0}=Z_{\cal O}{\cal O}. %
\eeq%
where the $C^{0}$ and ${\cal O}^{0}$ represent the unrenormalized
coefficient and operator.

Solving the RG equation for $C(\mu)$, we obtain %
\beq %
C(m_b)=C(m_W)~{\rm exp}\left[ \int_{g(m_W)}^{g(m_b)}dg
  \frac{\gamma(g)}{\beta(g)} \right].
\eeq %
This solution automatically sums large logarithms $\ln(m_W/m_b)$ to
all orders.

For the Sudakov form factor, the key point is that the anomalous
dimension contains a momentum dependent term. It is this
momentum-dependent anomalous dimension which distinguishes Sudakov
form factor from other physical quantities and dictates the
suppression of Sudakov form factor in the asymptotic limit. A
related anomalous dimension, cusp dimension had been known for a
long time \cite{Polyakov, KR}. A connection between the cusp
dimension and the anomalous dimension in Heavy Quark Effective
Theory (HQET) was pointed out in \cite{KR2}. As will be shown, this
cusp dimension is also closed related to the anomalous dimension of
soft-collinear effective theory. To some extent, the cusp dimension
is a fundamental quantity of QCD for interactions of soft gluons
with heavy-heavy, heavy-light, light-light quarks where heavy and
light represent the heavy and collinear quarks, respectively.

In this paper, we will study the Sudakov form factor in the
framework of SCET. Similar to \cite{Collins, Korchemsky}, we
consider the on-shell case. Because the Sudakov form factor had been
calculated long ago, we don't intend to provide a new and detailed
calculation. Our purpose is to look at the same topic from a point
of view inspired by the effective field theory. This view is not
totally new and most opinions had been implied in the previous
different methods. However, different considerations may involve
quite different technical details and physical interpretations. We
hope that our treatment of the Sudakov form factor in effective
theory can provide some useful insights. Our discussions will
concentrate on three aspects: factorization, evolution and physical
interpretation of the Sudakov form factor.

The traditional method uses a diagrammatic analysis to prove the
validity of factorization (or say, factorization theorem) to all
orders \cite{CSS}. A comparison of factorization within the SCET and
diagrammatic analysis is provided in \cite{B3}. In SCET, the proof
of factorization is replaced by integrating out the hard modes
(refers to the perturbative contributions in perturbative QCD) and
writing down all the possible low energy effective operators to
given orders of small expansion parameter $\lambda\sim m/Q$.  For
the separation of collinear gluons from the hard modes and soft
gluons from the collinear particles, the explicit soft and collinear
gauge invariance at the classical level simplify the discussions.

After integrating out the hard modes, it leads to a consistent RG
equation similar to Eq. (\ref{eq:RGE}). This is the result that
Sudakov form factor does not depend on choice of the renormalization
scale $\mu$. One intuitive understanding of the renormalization and
the RG equation is from the Wilson's renormalization group method
for critical phenomena in statistical physics \cite{WilsonRG}. Note
that idea of effective field theory originates from this method.
From \cite{WilsonRG}, the Sudakov form factor is a multi-scales
system rather than only two scales $Q, m$. It involves all the
intermediate scales between $Q$ and $m$. The procedure for
integrating out the intermediate momentum fluctuations scale by
scale form a cascade chain to give a RG equation and a
deamplification (suppression) effect. In \cite{Sterman}, it is
pointed out that each QCD evolution equation (such as DGLAP, BFKL
and Sudakov evolution equation) is associated with a cascade
mechanism represented by ladder diagram. We find a strong similarity
of the cascade mechanism in the renormalization group method and the
leading (double-)logarithmic approximation method. Based on this
understanding, we give an interpretation of the Sudakov form factor
from scale point of view.


This paper is organized as following: In sect. 2, we discuss the
Wilson lines and SCET in brief and then calculate the Sudakov form
factor in the framework of SCET. In sect. 3, we compare the
different approaches of Sudakov resummation and discuss the physical
interpretation of the Sudakov form factor. In sect. 4, the brief
discussions and conclusions are given.

\section{The Sudakov form factor in SCET}
\label{sec:SCET}

\subsection{Wilson lines and SCET}

One property of the SCET is that it involves different types of
Wilson line. In principle, the appearance of the Wilson line is
due to the local gauge invariance of QCD.  The QCD Lagrangian for
a massless quark field $\psi(x)$ is written as ${\cal
L}=\bar\psi(i\dslash)\psi$ where
$D_{\mu}=\partial_{\mu}-ig_sA_{\mu}=\partial_{\mu}-ig_sT^aA_{\mu}^a$.
The local gauge invariance permits us to write a formal form as%
\beq \label{eq:line}%
\psi(x)=W(x)~\psi^0(x), ~~~~~~~ W(x)=\exp\left( ig_s\int_{C}^x
dy^{\mu}A_{\mu}(y) \right).
\eeq %
where $C$ represents a path and ${\rm P}$ denotes path-ordering. It
should be noted that the Eq. (\ref{eq:line}) is a formal formulae.
Under the above transformation or say the field redefinition, all
effects of the gluon fields are included in a path-dependent phase
factor $W(x)$. The $\psi^0$ is the quark field with no interaction
with gluons and it satisfies the equation of motion for free quark
$i\parslash \psi^0=0$. The function of $W(x)$ is called Wilson line
which accumulates infinite gluons along a path. The path-dependent
phase factor of the Wilson line had been introduced for a long time.
A closed-path form of the Wilson line (called Wilson loop) is
proposed as a mechanism of quark confinement \cite{Wilson}.

In QCD, the infrared (IR) contributions are enhanced by IR
divergences when the virtual fields become on-shell. These on-shell
fields behave like classical particles and have an infinity numbers.
These IR particles may be analogous to the case of confined
particles. If the Wilson lines can be applicable to absorb a lot of
gluons, it will lead to a great simplification in theoretical
analysis. In SCET, which is a low energy effective theory of QCD to
describe the soft and collinear particles, we will see the
appearance of soft and collinear Wilson lines and these Wilson lines
are indispensable quantities.

It is convenient to use the light-cone coordinates to study the
processes with energetic light hadrons or jets. An arbitrary
four-vector $p^{\mu}$ is written as $p^{\mu}=(p^+, p^-,
p_{\bot})=(n_-\cdot p)\frac{n_+^{\mu}}{2}+ (n_+\cdot
p)\frac{n_-^{\mu}}{2}+p_{\bot}^{\mu}$ where
$n_+^{\mu}=(2,0,0_{\bot})$ and $n_-^{\mu}=(0,2,0_{\bot})$ are two
light-like vectors which satisfy $n_+^2=n_-^2=0$, and $n_+\cdot
n_-=2$. A four-component Dirac field $\psi$ can be decomposed into
two-component spinors $\xi$ and $\eta$ by
$\psi=\xi+\eta=\frac{\nslash_-\nslash_+}{4}\psi+
\frac{\nslash_+\nslash_-}{4}\psi$ with
$\nslash_-\xi=\nslash_+\eta=0$. The $\eta$ field is the heavy mode
need to integrated out from the effective theory. The momenta of the
collinear and soft particles are scaled as $p_c\sim Q(\lambda^2, 1,
\lambda)$ and $p_s\sim Q(\lambda, \lambda,\lambda)$ where
$\lambda=\lqcd/Q$.

In this study, we concern only the lowest order interaction of
$\lambda$ of the collinear and soft particles. Because we discuss
the on-shell Sudakov form factor, the ultrasoft particles will not
be considered. To simplify the illustration, we discuss a case that
collinear particles move close to $n_-$ direction. Other cases can
be given straightforwardly. The Lagrangian which describes the
interaction of the collinear quark with collinear gluons is written
as \cite{B2, BCDF}%
\beq \label{eq:cc}%
{\cal L}_{c}=\bar{\xi}\left [ ~in_-\cdot D_c
  +i\dslash_{c\bot}\frac{1}{~in_+\cdot D_c}~ i\dslash_{c\bot}
  \right ]\frac{\nslash_+}{2}~\xi.
\eeq %
where $D_c$ represent the covariant derivative for collinear
momentum regions. One property of the collinear Lagrangian is that
it is non-local which is different from other effective field
theories of QCD. The reason for this non-local interaction is that
the momentum component $p^-$ of collinear particles is at the same
order of the virtuality of the heavy mode.

For the interactions of collinear fields with soft gluons, the
momentum of the collinear particle does not retain its scaling when
a soft particle couples to it, $p_c+p_s\sim (\lambda, 1, \lambda)$.
The effective Lagrangian given in \cite{DG, Wei} can only be
interpreted as an intermediate theory. In \cite{B3}, it is proved
that the soft gluons decouple from the collinear quark or gluon in
the lowest order of $\lambda$. The
effects of soft gluons are included in the soft Wilson line %
\beq %
W_s(x)={\rm P~exp}\left ( ig_s\int_{-\infty}^x dt ~n_-\cdot
A_s(tn_-) \right ).
\eeq %
where the path-ordering ${\rm P}$ defined such that the the gluon
fields stand to the left for larger values of parameter $t$. The
soft Wilson line $W_s(x)$ describes the effect of infinite soft
gluons moving along the $n_-$ direction from $-\infty$ to point $x$.
For the collinear gluons, the case is different. We cannot decouple
the collinear gluons from the collinear quark in the same way as the
soft gluons. The thing we can do is to decouple collinear gluon
$n_+\cdot A_c$ from the denominator in Eq. (\ref{eq:cc}). This
can be expressed as %
\beq \label{eq:Wc}%
\frac{1}{~in_+\cdot D_c}=W_c\frac{1}{~in_+\cdot
 \partial}W_c^{\dagger}, ~~~~~W_c(x)={\rm P~exp}
 \left (ig_s\int_{-\infty}^x dt ~n_+\cdot A_c(tn_+) \right ).
\eeq %
The Eq. (\ref{eq:Wc}) means that when we integrate out the hard mode
(it refers to heavy degrees of freedom in the effective field
theory), the $n_+\cdot A_c$  collinear gluons can be grouped into a
Wilson line along the $n_+$ direction. In other words, the coupling
of collinear gluons to the hard mode is equivalent to the coupling
to a Wilson line. Another explanation of the above soft and
collinear Wilson lines is that the longitudinal polarized gluon
($A_{\pm}$) is unphysical thus it can be gauged into a phase factor
due to gauge invariance \cite{CSS}.

The SCET has a remnant gauge invariance under the collinear and soft
transformations which do not change momentum fluctuations of the
collinear and soft particles. The collinear and soft gauge
transformations $U(x)$ are constrained by momentum regions $\partial
U_c(x)\sim (\lambda^2, 1, \lambda)$, $\partial U_s(x)\sim (\lambda,
\lambda, \lambda)$. The collinear fields transform in the usual way
under the collinear gauge transformation as in the classical theory.
The Lagrangian in Eq. (\ref{eq:cc}) is invariant under the collinear
gauge invariance. The collinear fields do not transform under the
soft gauge transformation because the coupling of soft particles
lead to off-shellness of collinear particles. The soft fields also
do not transform under the collinear gauge transformation.

\subsection{The factorization of the Sudakov form factor}

The asymptotic quark form factor provides a simple example to
discuss the Sudakov resummation. In \cite{Wei}, we proved that SCET
reproduces all the IR physics of the full theory of QCD in the quark
form factor at one-loop order. In the Appendix,  a more detailed
calculation than in \cite{Wei} is presented for reference. Here, we
discuss the resummation of Sudakov-logs to all orders in SCET. The
notations are given as same as in \cite{Wei}. We consider an
electromagnetic form factor of a quark given by $\langle
q_B|\bar\psi_B\Gamma\psi_A |q_A \rangle=\bar u(p_B)\gamma u(p_A)
F(Q^2)$ with $\Gamma=\gamma_{\mu}$ and study a case that the the
initial and final quarks $q_A$ and $q_B$ are both massless and
on-shell. Their momenta are chosen as $p_A=(Q, 0, 0_{\bot}),
~p_B=(0, Q, 0_{\bot})$ and $q^2=(p_B-p_A)^2=-Q^2$ where $Q$ is a
large energy scale.

\begin{figure}
\includegraphics[scale=1.2]{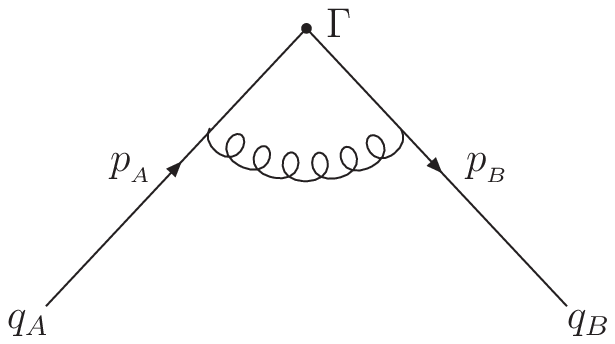}
\caption{The one-loop vertex correction to the quark form factor.}
\label{fig:vertex}
\end{figure}

Let us consider the current operator
$V_{\mu}=\bar\psi_B\gamma_{\mu}\psi_A$. In the full theory, the
vector current does not require renormalization because of current
conservation. The matrix element  $\langle
q_B|\bar\psi_B\gamma_{\mu}\psi_A |q_A \rangle$ has no ultrasoft (UV)
divergence\footnote{The UV divergences in vertex corrections are
cancelled by the quark field renormalization.} and the form factor
$F(Q^2)$ is independent of the renormalization scale $\mu$. But
$F(Q^2)$ contains logarithms $-\as {\rm ln}^2\frac{Q^2}{\delta^2}$
in one-loop vertex correction depicted in Fig. {\ref{fig:vertex}}
where $\delta$ is a low energy scale. The appearance of the large
logarithms is due to the existence of separate scales in a system.
It means the breakdown of the usual perturbation theory.

A way to disentangle the scales is to substitute the full theory
with simpler but equivalent effective theories by systematically
integrating out the heavy degrees of freedom scale by scale.  The
basic idea of summing large logarithms in the effective theory had
been explained in the Introduction. Now we use the SCET to separate
the scales in the Sudakov form factor.

The hard region with  virtual momenta of $k\sim Q(1,1,1)$ is the
hard mode which contributes to the Wilson coefficient. An infinite
collinear gluons couple to the hard loops and the collinear quark
moves in another direction have momentum virtuality of $Q^2$ without
suppression in leading order of $\lambda$. Because of gauge
invariance, it is convenient to use the explicit gauge-invariant
quantities. The collinear quark $\xi(x)$ transforms to
$U_c(x)\xi(x)$ under the collinear gauge transformation. The
collinear Wilson line transforms as $W_c(x)\to U_c(x)W_c(x)$. A
gauge invariant combination of them is is a gauge singlet
$W_c^{\dagger}\xi$ under collinear gauge transformations. This gauge
singlet operator includes the interactions of collinear gluons with
the hard loops and the collinear quark moves in another direction.

The couplings of soft gluons to the collinear filed lead to
off-shellness of $Q^2\lambda\gg Q^2 \lambda^2$. Integrating out this
off-shell modes gives soft Wilson line $W_s$. After this, the soft
gluons decouple from the collinear fields. Under the soft gauge
transformations $U_s(x)$, the collinear fields is unchanged and the
soft Wilson lines transforms as $W_s(n_-)\to U_s W_s(n_-)$,
$W_s(n_+)\to U_s W_s(n_+)$ where $W_s(n)$ represents the collinear
Wilson line along the $n$ direction. The combination of
$W_s(n_-)^{\dagger}W_s(n_+)$ is invariant under the soft gauge
transformations.

The above discussions give a gauge invariant expression for the
current operator in SCET as %
\beq \label{eq:vnu}%
\bar\xi_{n_-}W_c(n_+)W_c^{\dagger}(n_-)\xi_{n_+}(\mu)
 W_s(n_-)^{\dagger}W_s(n_+).
\eeq %
where $\xi_{n_+}$ and $W_c(n_-)$ represent the collinear fields move
along the $n_-$ direction. Similar interpretations for the other
operators are implied. The effective operator includes all the
low-energy dynamics but no high-energy dynamics. The hard mode with
virtualities of $k\sim Q(1,1,1)$ needs to be included when we match
the full theory onto the effective theory since the prediction of
the two methods must be equal to a given order of $\lambda$. The
matching of the current operator gives %
\beq %
V_{\Gamma}=\int ds ds' \tilde{C}(s,s',\mu)\left[\bar
\xi_{n_-}W_c(n_+)\right](sn_+)\Gamma
\left[W_c(n_-)\xi_{n_+}\right](s'n_-) W_s(n_-)^{\dagger}W_s(n_+),
\eeq %
The $\tilde{C}(s,s',\mu)$ is position space Wilson coefficient which
depends on the position of the collinear field. The appearance of
integral over $s$ is due to that the momenta $p_A^+$, $p_B^-$ are at
order of $Q$.

The matrix element $\langle q_B|\bar\psi_B\gamma_{\mu}\psi_A |q_A
\rangle$ then becomes %
\beq %
\langle q_B|\bar\psi_B\gamma_{\mu}\psi_A |q_A \rangle&=&\gamma_{\mu}
 \int ds ds' \tilde{C}(s,s',\mu)
 \left\langle q_B\left|\left[\bar \xi_{n_-}W_c(n_+)\right]
  (sn_+)\right|0\right\rangle \non \\
&& ~~\times\left\langle 0\left|\left[W_c(n_-)\xi_{n_+}\right]
  (s'n_-)\right|q_A \right\rangle
 \langle 0|W_s(n_-)^{\dagger}W_s(n_+)|0\rangle \non \\
&=& C(n_+\cdot p_B, n_-\cdot p_A, \mu)\left\langle
q_B\left|\left[\bar \xi_{n_-}W_c(n_+)\right]
  \right|0\right\rangle \non \\
&& ~~\times\left\langle 0\left|\left[W_c(n_-)\xi_{n_+}\right]
  \right|q_A \right\rangle
 \langle 0|W_s(n_-)^{\dagger}W_s(n_+)|0\rangle
\eeq %
In the above equation, we have used the translation invariance
$\phi(a)=e^{ia\cdot P}\phi(0)e^{-ia\cdot P}$. The $C(n_+\cdot p_B,
n_-\cdot p_A, \mu)$ is the momentum space Wilson coefficient defined
by %
\beq %
C(n_+\cdot p_B, n_-\cdot p_A, \mu)=\int ds ds' e^{isn_+\cdot
p_B}e^{-isn_-\cdot p_A}~\tilde{C}(s,s',\mu),
\eeq %
In SCET, there is remnant Lorentz invariance called by
reparameterization invariance. One class is the longitudinal boosts
$n_+\to \alpha n_+$, $n_-\to \alpha^{-1} n_-$. The Lorentz
invariance constraints the $C(n_+\cdot p_B, n_-\cdot p_A, \mu)$ can
only depend on $(n_+\cdot p_B) (n_-\cdot p_A)=Q^2$. The
dimensionless hard Wilson coefficient make us to simplify
$C(n_+\cdot p_B, n_-\cdot p_A, \mu)=C(Q/\mu)$.

Thus, we obtain a final  explicit factorized form for the form factor
$F(Q^2)$ as %
\beq %
F(Q^2)=C(\mu) ~ J_A ~ J_B~ S.
\eeq %
where $J_A$, $J_B$ and $S$ are defined by %
\beq \label{eq:factorization}%
J_A &\equiv& \langle 0|{\rm P ~exp} \left (ig_s\int_0^{+\infty}
 dt ~n_-\cdot A_c(tn_-) \right )\xi_{n_+}|q_A\rangle; \non \\
J_B &\equiv& \langle q_B|\bar\xi_{n_-}~ {\rm P ~exp} \left (
 ig_s\int_{-\infty}^0 dt ~n_+\cdot A_c(tn_-) \right )|0\rangle; \\
S~&\equiv& \langle 0|{\rm P ~exp} \left (
 ig_s\int_0^{+\infty} dt ~n_-\cdot A_s(tn_-) \right )
 {\rm exp}\left ( ig\int_{-\infty}^0 dt
 ~n_+\cdot A_s(tn_+) \right )|0\rangle.\non
\eeq %
The above factorization formulae is consistent with the result given
in \cite{Collins, Korchemsky}.

\begin{figure}
\includegraphics[scale=1.2]{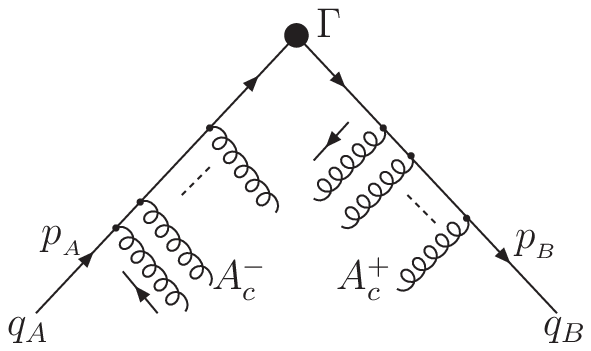}
\caption{The collinear gluons contribute to the collinear Wilson
lines. The $A_c^+$ and $A_c^-$ represent the gluons collinear to
quarks $q_A$ and $q_B$ respectively.} \label{fig:Wc}
\end{figure}

\begin{figure}
\includegraphics[scale=1.2]{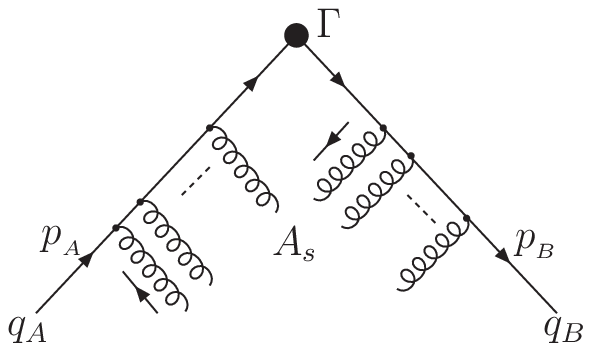}
\caption{The soft gluons contribute to the soft Wilson lines.}
\label{fig:Ws}
\end{figure}

Expanding the path-ordered exponential of Wislon line $W$ in
orders of the coupling constant gives the Feynman
rules in momentum space %
\beq %
{\rm P ~exp} \left (
 ig_s\int_0^{+\infty} dt ~n\cdot A(tn) \right )=1+{\rm P}
 \sum\limits_{i=1}^\infty  \prod\limits_{j=1}^i
 \int \frac{d^4 k_j}{(2\pi)^4}ig_s n\cdot \tilde{A}(k_j)
 \frac{i}{n\cdot \sum_{j=1}^i k_j}.
\eeq %
where $\tilde{A}(k_j)$ are Fourier conjugated field of gluon
$A(tn)$.

The soft and collinear Wilson lines are plotted in Fig.
\ref{fig:Wc} and Fig. \ref{fig:Ws}. The jet-A function is defined
as a matrix element of the quark field $A$ with a path-ordered
Wilson line which accumulates an infinite gluons collinear to A
from point 0 (the hard scattering point) to $+\infty$ along the
$n_-$ direction. The jet-B function is defined as a matrix element
of the quark field $B$ with a path-ordered Wilson line which an
infinite collinear-to-B gluons go from $-\infty$ to the point $0$
along the $n_-$ direction. The soft function $S$ is independent of
the quark flavor and spin. This is due to an additional
spin-flavor symmetry for the interactions of collinear quark with
soft gluons in SCET. The $S$ is defined as the vacuum expectation
value of soft Wilson lines. The soft gluons go from $-\infty$ to
point 0 along the $n_+$ direction and then from point 0 to
$+\infty$ along the $n_-$ direction. Note that the Wilson line is
unitary, i.e., $W^{\dagger}W=1$. The path-ordering in the Wilson
line is similar to the time-ordering in the conventional quantum
field theory and the parameter $t$ acts as time.

\subsection{The summation of Sudakov-logs in SCET}

The calculation of the Wilson coefficient $C(\mu)$ is performed by
a matching procedure from the full theory onto the effective
theory. Because the origin of the Wilson coefficient $C(\mu)$ is
insensitive to the detail of IR physics, one is free to choose any
infrared regularization method and the external states. For the
quark form factor, the most convenient way is to use the
dimensional regularization method and to perform the matching on
mass shell of the massless quark. In the dimensional
regularization, all the loop corrections to the long-distance
functions $J_A$, $J_B$ and $S$ vanish because there is no scale
parameter in the loop integral. The only remained integral is
coming from contribution of the hard part which the momentum of
the virtual quark $k\sim Q (1,1,1)$. The one-loop contribution to
the quark form factor in $d=4-2\epsilon$ dimension
is%
\beq \label{eq:hard}%
I&=&-ig_s^2 C_F\mu^{\prime 2\epsilon}\int\frac{d^d k}
  {(2\pi)^d} \frac {\gamma_{\rho}(\kslash+\pslash_B)
  \gamma_{\mu}(\kslash+\pslash_A) \gamma^{\rho}}
  {[k^2+2p_A\cdot k+i\epsilon][k^2+2p_B\cdot k+i\epsilon]
  [k^2+i\epsilon]} \non \\
 &=&\frac{\as}{4\pi}C_F \gamma_{\mu}\left[ -\frac{2}{\epsilon^2}
  -\frac{3+2{\rm ln}\frac{\mu^2}{Q^2}}{\epsilon}
 -{\rm ln}^2\frac{\mu^2}{Q^2}-3{\rm ln}\frac{\mu^2}{Q^2}
 -8+\frac{\pi^2}{6} \right].
\eeq %
where $\mu^2=4\pi \mu^{\prime 2}{\rm e}^{-\gamma_E}$ is the scale
defined in the ${\rm \overline{MS}}$ scheme and $\gamma_E$ is the
Euler constant.

The one-loop correction is divergent and requires renormalization.
This is different from the result in the full theory. We need not
care about whether the divergence is of UV or IR type in the
dimensional regularization when doing renormalization, although the
divergences are IR in origin. The renormalization in the effective
field theory can be done by the standard counter-term method. The
coefficient $C$ is regarded as coupling constant. One way to perform
the renormalization is done by redefining the coefficient $C$ while
leaving the operators in SCET
unrenormalized as follows %
\beq \label{eq:cr}%
C^0=Z_C C, ~~~~~~~Z_C=1-\frac{\as}{4\pi}C_F \left[
  \frac{2}{\epsilon^2}+\frac{3+2{\rm ln}\frac{\mu^2}{Q^2}}
  {\epsilon}\right].
\eeq %
The $C^0$ represents the bare coupling constant.  The
renormalization constant $Z_C$ contains momentum $Q$-dependent
counter-term which will contribute to a momentum-dependent anomalous
dimension. There is a $\frac{1}{\epsilon^2}$ double-poles in one
loop order of the $Z_C$. This is not familiar in the standard
perturbation theory. The appearance of the double-poles is due to
the overlap of soft and collinear divergences in the effective
theory when the low energy scale approaches 0. The $Z_C$ does not
depend on the low-energy scale. The renormalized
matching coefficient $C(\mu)$ up to one-loop order is %
\beq %
C(\mu)=1+\frac{\as}{4\pi}C_F \left[
 -{\rm ln}^2\frac{\mu^2}{Q^2}-3{\rm ln}\frac{\mu^2}{Q^2}
 -8+\frac{\pi^2}{6} \right].
\eeq %

The renormalization in SCET can be done in a different but
equivalent way. Another method is the conventional operator
renormalization which renormalizes the operators rather than the
coefficients. For this method, we need to know the exact UV
divergences of the operators in SCET. We also provide this
renormalization method for reference. From the results of
Appendix, the renormalization constants for $J_A$, $J_B$ and $S$ are %
\beq \label{eq:ct}%
Z_A:~~~~~ && \frac{\as}{4\pi}C_F \left[\frac{2}{\epsilon^2}
   +\frac{2+2{\rm ln}\frac{\mu^2}{Q \delta}}{\epsilon}+\frac{-1}
   {2\epsilon}  \right], \non \\
Z_B:~~~~~ && \frac{\as}{4\pi}C_F \left[\frac{2}{\epsilon^2}
   +\frac{2+2{\rm ln}\frac{\mu^2}{Q \delta}}{\epsilon}+\frac{-1}
   {2\epsilon}  \right], \non \\
Z_S:~~~~~ && \frac{\as}{4\pi}C_F \left[-\frac{2}{\epsilon^2}
   -\frac{2{\rm ln}\frac{\mu^2}{\delta^2}}{\epsilon} \right], \non\\
Total:~~~  && \frac{\as}{4\pi}C_F \left[\frac{2}{\epsilon^2}
  +\frac{3+2{\rm ln}\frac{\mu^2}{Q^2}}{\epsilon}\right],
\eeq %
where the $\frac{\as}{4\pi}C_F\frac{-1}{2\epsilon}$ terms for
$J_A$ and $J_B$ are coming from the quark field $\xi$
renormalization. The total counter-term $Z_T$ can be
obtained from  Eq. (\ref{eq:ct}) as %
\beq \label{eq:zt} %
Z_T=\frac{\as}{4\pi}C_F \left[\frac{2}{\epsilon^2}
  +\frac{3+2{\rm ln}\frac{\mu^2}{Q^2}}{\epsilon}\right].
\eeq %
$Z_T$ is the negative of the relevant term of $Z_C$ in Eq.
(\ref{eq:cr}). There is a general relation of the renormalization
constants in the operator renormalization and the coefficient
renormalization: $Z_T=Z_C^{-1}$ \cite{Buras}.

In Eq. (\ref{eq:cr}), the bare coupling constant $C^0$ is
$\mu$-independent. Using the relation $\mu\frac{dC^0}{d\mu}=0$, we
obtain a renormalization group equation for $C(\mu)$ as %
\beq \label{eq:rge} %
\mu\frac{d}{d\mu}C(\mu)=\left(\mu\frac{\partial}{\partial \mu}
  +\beta(g)\frac{\partial}{\partial g} \right)C(\mu)=\gamma_C
  C(\mu).
\eeq %
Here, the anomalous dimension $\gamma_C$ is defined by %
\beq %
\gamma_C=-\frac{1}{Z_C}~\frac{d Z_C}{d{\rm ln}\mu}
  =\frac{1}{Z_T}~\frac{d Z_T}{d{\rm ln}\mu},
\eeq %
The direct way to calculate the anomalous dimension is through the
$1/\epsilon$-pole terms of the renormalization constant. From the
result in Eq. (\ref{eq:cr}), the anomalous dimension
$\gamma_C$ is obtained  up to $\as$ order as %
\beq %
\gamma_C=2\as\frac{\partial Z_{C,1}}{\partial \as}=
  -\frac{\as}{2\pi}C_F \left(3+2{\rm
  ln}\frac{\mu^2}{Q^2}\right).
\eeq %
where $Z_{C,1}$ is the  $1/\epsilon$-pole coefficient of $Z_C$. Note
that in the derivation of the above evolution equation we used the
non-renormalization property for the collinear and soft fields. One
property of the anomalous dimension $\gamma_C$ is its Q-dependence.
Another thing is that the anomalous dimension in leading logarithmic
approximation (here, the leading logarithmic approximation refers to
the leading double-logarithmic approximation) is positive, i.e.,
$\gamma_C=\frac{\as}{\pi}C_F{\rm ln}\frac{Q^2}{\mu^2} >0$ for
$\mu\ll Q$. That means $C(\mu)$ is a decreasing function as $Q$
increases or $\mu$ decreases. This provides an explanation of the
suppression of the Sudakov form factor.

The solution of the renormalization group equation like Eq.
(\ref{eq:rge}) is an exponential form in general%
\beq \label{eq:sr}%
C(\mu_0)&=&C(Q)~{\rm exp}\left[ \int_{{\rm ln }Q}^{{\rm ln}\mu_0}
  d({\rm ln}\mu) \gamma_C(g(\mu)) \right]  \non \\
  &=&C(Q)~{\rm exp}\left[ \int_{g(Q)}^{g(\mu_0)}dg
  \frac{\gamma_C(g)}{\beta(g)} \right].
\eeq %
For our case, the one-loop calculations give $C(Q)=
1+\frac{\as}{4\pi}C_F \left[ -8+\frac{\pi^2}{6} \right]$ and
$\gamma_C= -\frac{\as}{2\pi}C_F \left(3+2{\rm
ln}\frac{\mu^2}{Q^2}\right)$. The scale $\mu_0$ is a low energy
scale but be chosen to guarantee the smallness of the coupling
constant $\as$.

The solution of the renormalization group equation sums the series
of large logarithms $\as^n{\rm ln}^m Q^2(m\le 2n)$ to all orders
automatically. We define the leading-log (LL) and
next-to-leading-log (NLL) approximations for summation as %
\beq %
{\rm LL~~}:~~~~~ &&  \sum\limits_n \as^n ~({\rm ln}Q^2)^{2n}; \non \\ %
{\rm NLL}:~~~~~  &&  \sum\limits_n \as^n ~({\rm ln}Q^2)^{2n-1}.
\eeq %
The LL and NLL approximations are valid if the relations below are
satisfied %
\beq %
\as\ll \as\ln Q^2\ll \as\ln^2 Q^2\sim {\cal O}(1).
\eeq %

At first sight, the coefficient $C(\mu_0)$ has no relation with
the Sudakov form factor. The meaning of the coefficient $C(\mu_0)$
will be clear after we discuss a simple case of the solution of
Eq. (\ref{eq:rge}). In the LL approximation, %
\beq \label{eq:rcLL}%
C(Q)=1, ~~~~~~ %
\gamma_C^{\rm LL}= -2\frac{\as}{\pi}C_F \ln\frac{\mu}{Q}.
\eeq %
We consider a case that the coupling constant $g_s$ is frozen at a
finite value, or say the running effect is neglected. From Eq.
(\ref{eq:sr}), we obtain a solution for this simple case as %
\beq %
C(\mu_0)={\rm exp}\left[ -\frac{\as}{4\pi}C_F {\rm
  ln}^2 \frac{Q^2}{\mu_0^2}\right].
\eeq %
This reminds us the familiar result $F={\rm exp}\left[
 -\frac{g^2}{16\pi^2} {\rm ln}^2 \frac{Q^2}{m^2}\right]$ in QED
when $C_F=1$ and $\mu_0=m$. We can say that: the Wilson coefficient
$C(\mu_0)$ is the perturbative part of the usual Sudakov form
factor. When the quark confinement is ignored, $C(\mu_0)$ is equal
to $F(Q^2)$. The exponentiation of the Sudakov form factor is
explained by the solution of a renormalization group equation. The
mechanism for the exponentiation of Sudakov form factor is the same
as other physical quantities which satisfy the RG evolution
equations.

The effect of the running of the coupling constant can be included
straightforwardly by using
$\as(\mu)=\frac{4\pi}{\beta_0~\ln(\mu^2/\Lambda^2)}$ and
$\beta(g)=-\beta_0 g\frac{\as}{4\pi}$ in LL approximation. The
solution of $C(\mu_0)$ is %
\beq %
C(\mu_0)=\exp \left\{ -4\frac{C_F}{\beta_0} \left[
  \ln \frac{Q}{\Lambda} \ln\frac{\ln Q/\Lambda}
  {\ln \mu_0/\Lambda}-\ln \frac{Q}{\mu_0} \right]\right\}.
\eeq %
or written in another form %
\beq \label{eq:LL}%
C(\mu_0)=\exp \left \{ \frac{8\pi C_F}{\beta_0^2\as(Q)}
 \left[ 1-\frac{1}{z}-\ln z \right]\right\}.
\eeq %
where $z=\frac{\as(\mu_0)}{\as(Q)}$ is defined in \cite{B2}.

In NLL approximation, the anomalous dimension contains one-loop
order as well as two-loops order corrections. We will not do an
explicit two-loops calculation but use the results provided
in \cite{KR, Collins}. In NLL, we the anomalous dimension is written
as %
\beq \label{eq:rcNLL}%
\gamma_C^{\rm NLL}=-\frac{\as}{4\pi}C_F 6-(\frac{\as}{\pi})^2
 C_F B~\ln \frac{\mu}{Q},
\eeq %
where the coefficient $B$ will be determined by other methods.

The momentum-dependent anomalous dimension has been calculated as
a cusp dimension up to two-loops order as \cite{KR, Collins}%
\beq \label{eq:cusp1}%
\Gamma=-\left\{ 2\frac{\as}{\pi}C_F +\left[
 \left(\frac{67}{18}-\frac{\pi^2}{6}\right)C_A-
 \frac{10}{9}n_f T_F \right]\left(\frac{\as}{\pi}\right)^2
 C_F\right\}\ln\frac{\mu}{Q},
\eeq %
where $n_f$ is the number of quark flavors and the SU(3)$_C$ group
constants are: $T_F=\frac{1}{2}$, $C_F=\frac{4}{3}$ and $C_A=3$.

From the comparison of Eqs.(\ref{eq:rcLL}, \ref{eq:rcNLL}) with
Eq. (\ref{eq:cusp1}), we obtain the coefficient $B$ as %
\beq %
B=\left(\frac{67}{18}-\frac{\pi^2}{6}\right)C_A-
 \frac{10}{9}n_f T_F .
\eeq %

In order to calculate the Sudakov form factor to NLL, we need the
function $\beta(g)$ and coupling constant $\as$ to next-to-leading
order \cite{Buras}, %
\beq %
\beta(g)&=&-\beta_0 g \frac{\as}{4\pi}\left(1+\frac{\beta_1}
  {\beta_0}\frac{\as}{4\pi}\right), \non \\
\as(\mu)&=&\frac{4\pi}{\beta_0~ \ln (\mu^2/\Lambda^2)}
  \left( 1-\frac{\beta_1}{\beta_0^2}\frac{\ln~\ln(\mu^2/\Lambda^2)}
  {\ln(\mu^2/\Lambda^2)} \right), \non \\
\ln\frac{\mu}{Q}&=&\frac{2\pi}{\beta_0}\left[
 \frac{1}{\as(\mu)}-\frac{1}{\as(Q)}+
 \frac{\beta_1}{4\pi\beta_0}\ln\frac{\as(\mu)}{\as(Q)} \right].
\eeq %
where $\beta_0=\frac{(11N_c-2n_f)}{3}$ and
$\beta_1=\frac{34}{3}N_c^2-\frac{10}{3}N_c n_f-2C_F n_f$ with
$N_c=3$.

The final formula for coefficient $C(\mu_0)$ up to NLL approximation
is %
\beq %
C(\mu_0)=e^{S(Q,\mu_0)}.
\eeq %
where we have taken approximation $C(Q)=1$. The factor $S(Q,\mu_0)$ is%
\beq %
&&~S=S^{LL}+S^{NLL},~~~~~~~~~ %
S^{LL}=\frac{8\pi C_F}{\beta_0^2\as(Q)}
 \left[ 1-\frac{1}{z}-\ln z \right], \non \\
&&S^{NLL}=\frac{3C_F}{\beta_0}\ln z
 +\frac{4C_F}{\beta_0^2}B(1-z+\ln z)
 +\frac{2C_F\beta_1}{\beta_0^3}(z-1-\ln z+\frac{1}{2}\ln^2 z).
\eeq %
The term $S^{LL}(Q,\mu_0)$ is the result of the LL approximation
which had been given in Eq. (\ref{eq:LL}). The NLL term
$S^{NLL}(Q,\mu_0)$ is suppressed by a logarithm $\ln Q^2$ compared
to the leading factor $S^{LL}(Q,\mu_0)$.

\begin{figure}
\includegraphics[scale=1.2]{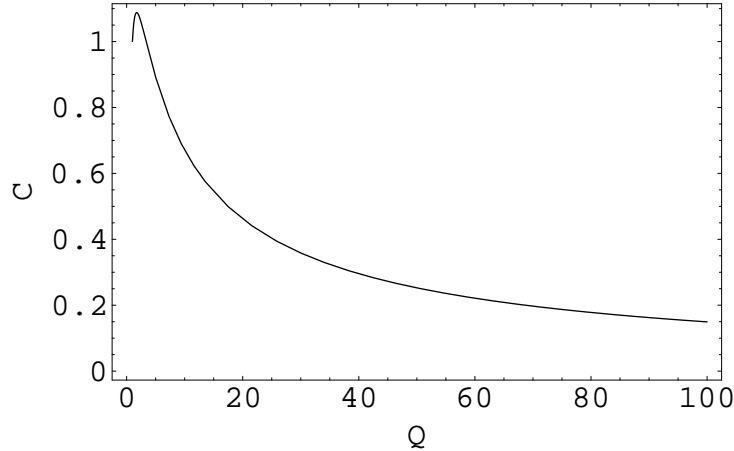}
\caption{The Q-dependence of Sudakov form factor $e^{S(Q,\mu_0)}$
including both the LL and NLL contributions. The variables Q are
given in units of GeV.} \label{fig:Sudakov}
\end{figure}

\begin{figure}
\includegraphics[scale=1.2]{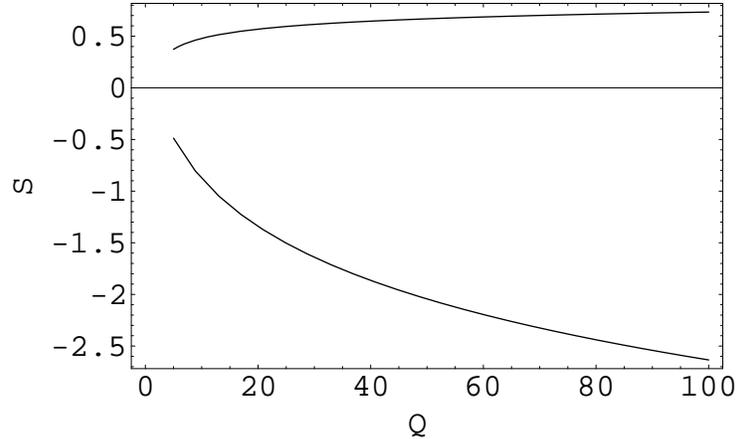}
\caption{The Q-dependence of factors $S^{LL}$ and $S^{NLL}$. The
upper curve is  plotted for $S^{NLL}$ and the lower one for
$S^{LL}$.} \label{fig:s0s1}
\end{figure}

\begin{figure}
\includegraphics[scale=1.2]{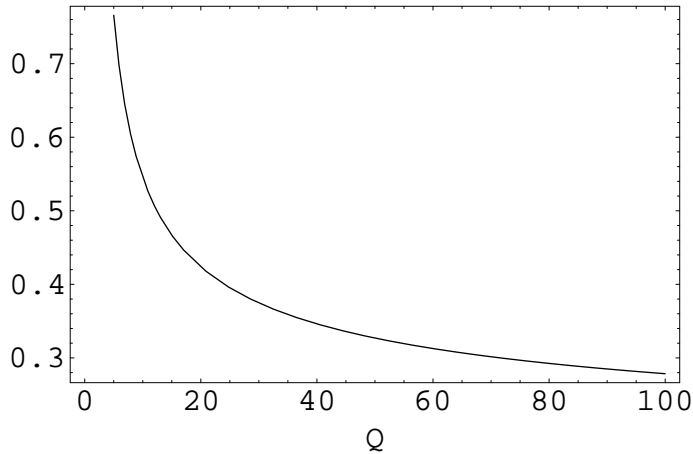}
\caption{The ratio of $-S^{NLL}/S^{LL}$ vs energy scale $Q$.}
\label{fig:ratio}
\end{figure}

The coefficient $C(\mu_0)$, or say the Sudakov form factor
$e^{S(Q,\mu_0)}$ is a decreasing function of $Q$. Fig.
\ref{fig:Sudakov} shows that Sudakov form factor $e^{S(Q,\mu_0)}$
damps as $Q$ increases. The suppression of the Sudakov form factor
is due to negative value of the LL factor $S^{LL}(Q,\mu_0)$.
However, the NLL factor $S^{NLL}(Q,\mu_0)$ is positive and has a
destructive effect on the suppression of the LL result. Fig.
\ref{fig:s0s1} plots the $Q$-dependence of factors $S^{LL}$ and
$S^{NLL}$. In order to ensure the convergence of summation series,
the NLL contribution should be much smaller than the leading one. We
check this by using the ratio $-S^{NLL}/S^{LL}$. From Fig.
\ref{fig:ratio}, the ratio $-S^{NLL}/S^{LL}$ becomes much smaller
than 1 when $Q$ is very large. The parameters are chosen as: the QCD
scale $\Lambda=0.25\GeV$; the quark flavor number $n_f=4$ and the
scale $\mu_0=1.0\GeV$.

\subsection{The interpretation of $F(Q^2)$}

In the discussions below, we will neglect the confinement effect and
set $C(\mu_0)=F(Q^2)$. Because the Sudakov form factor is a
dimensionless quantity, the naive expectation from tree level
consideration is that it is a constant: $F=1$. The radiative
corrections change it to dependent on energy $F(Q^2)=F(Q^2/m^2)$.
Because the radiative corrections is perturbative, one may expect
that the deviation of $F(Q^2)$ from $1$ is of order of $\alpha_s$
and thus small. The result from resummation tells us that the naive
thinking is wrong. The $F(Q^2)$ damps fast to zero when $Q$ is large
enough. One explanation of this non-trivial result is that the large
logarithms modify the perturbative series and the sum of them to all
orders is the correct result. Here, we give another explanation from
scale point of view: the Sudakov form factor is a multi-scales
system, the intermediate scales are all important and the sum of
contributions from all scales gives the Sudakov form factor.

The scales in the Sudakov form factor are $Q, ~m$ and the
intermediate scales between them. The crucial feature for the
intermediate region is the absence of characteristic energy scales.
Because of this feature, we can apply the Wilson's renormalization
group method \cite{WilsonRG}. The principle (or say assumption)
behind this method is that the many energy scales are locally
coupled. The dynamics associated with each energy scale can be
interpreted as a superposition in scale space. It means: the high
momentum fluctuations at $Q$ does not couple importantly to the low
momentum fluctuations at $m$, the coupling of momentum fluctuations
at scale $\mu$ to the fluctuations at scale $\mu/10$ is weaker than
to scale $\mu/2$. The result of the local property is a cascade
mechanism: the fluctuation of $Q$ influences fluctuation of $Q/2$;
the fluctuation of $Q/2$ influences fluctuation of $Q/4$; etc. until
to the low momentum fluctuations of $m$. The treatment of
multi-scales problem can be done by integrating out the fluctuations
scale by scale: first integrate out fluctuation of $Q$ and obtain
$C(Q)$, then integrate out fluctuations between $Q$ and $Q/2$ and
obtain $C(Q)$; etc. at the end we obtain $C(m)=F(Q^2)$. The cascade
chain can also be performed continuously from $\mu\to \mu-d\mu$.
This gives the evolution equation of Eq. (\ref{eq:rge}). The
solution of the evolution equation from high to low energy scales
gives the Sudakov form factor.

In SCET, the cascade mechanism is realized as: First, we integrate
out the momentum fluctuations at scale $\mu=Q$ and obtain an
effective current $V_{QCD}=C(Q)V_{SCET}(Q)$; Second, we integrate
out the intermediate scales step by step to the low energy which
corresponds to solve the evolution equation of Eq. (\ref{eq:rge});
At last we obtain $V_{QCD}=C(m)V_{SCET}(m)$, the Sudakov form factor
is included in $C(m)$.

Because there is no characteristic scale in the intermediate
regions, the similar effects should occur for each step of the
cascade chain. It leads to one feature of the cascade mechanism: the
existence of amplification or deamplification as cascade develops.
Whether the effect is amplification or deamplification depends on
the sign of dimension function in the evolution equation. In other
words, if the influence of fluctuations of scale $\mu$ on
fluctuations of $\mu-d\mu$ is negative, the deamplification occurs.
The Sudakov form factor belongs to the deamplification effect. The
larger space for the cascade developing when $Q$ increases, the
higher the suppression is. As have been discussed, it is the cusp
dimension dimension that determines the suppression of the Sudakov
form factor.

The above picture illustrates the idea behind the effective field
theory. It explains why the factorization in SCET are different from
the diagrammatic analysis. For the technical calculations, the
dimensional regularization method and the conventional
renormalization procedure are convenient and useful in perturbation
theory. Note that the application of the effective field theory does
not restricted in the perturbation theory, one example is the chiral
perturbation theory.

\section{Comparison with other methods of Sudakov resummation}

In the last section, we have discussed the Sudakov resummation in
the soft-collinear effective theory by utilizing the renormalization
group method. The Sudakov form factor had been got extensive studies
in the literatures. Here, we want to compare our approach with other
methods. However, the approaches\footnote{some approaches are
related with each other.} about Sudakov resummation appeared in
literatures  are too much to be considered fully. We choose three
methods for discussion: the leading double-logarithmic approximation
method, the Wilson loop method and the CSS method. We will not
concern the technical details but the main concepts.

\subsection{The leading double-logarithmic approximation method}

\begin{figure}
\includegraphics[scale=1.3]{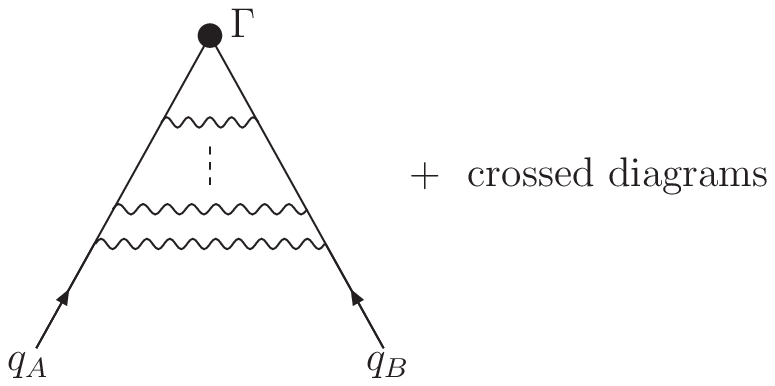}
\caption{The ladder graphs of quark form factor in QED in the
leading double-logarithmic approximation.} \label{fig:ladder}
\end{figure}

The explicit calculation of the Sudakov form factor order by order
is instructive to understand the relation between the RG method and
the Feynman diagram method. The calculation of the Sudakov form
factor beyond the $\as$ order in QCD is complicate due to the
self-interactions of gluon fields \cite{CPQ}. We will discuss the
case of QED at first to obtain some insights. In Feynman gauge, the
leading contributions (in leading double-logarithmic approximation)
to the vertex correction come from the ladder and crossed-ladder
graphs which is plotted in Fig. \ref{fig:ladder}. About the method
of the leading double-logarithmic approximation\footnote{Sudakov is
the first to apply this method \cite{Sudakov}. However, we are
failed to find his paper in our place, so we don't know the detail
of his treatment.}, we refer to recent papers of \cite{FLMM, ET}.

The mechanism for appearance of the double-logarithms is: an
energetic quark emits a soft and quasi-collinear photon and then the
photon is absorbed by another collinear quark. If the quark emits a
pure collinear or soft photon, there is only single logarithm. When
the soft and collinear regions overlap, the double-logarithms
appear. The extension of the leading order mechanism to all orders
can be pictured as a cascade chain $qq\to qq\to qq\to...$ by
infrared photon exchange which contributes to the ladder graphs in
QED. The quark scattering satisfies a local property that the quark
scattering of a given level does not depend on the details of the
scattering at a deeper level. The simplest way to express the
cascade chain is through an infrared evolution equation%
\beq \label{eq:LLA}%
F(\frac{Q^2}{\mu^2})=1-\int_{\mu^2}^{Q^2}\frac{\alpha(k_{\bot})}{2\pi}
 \frac{dk_{\bot}^2}{k_{\bot}^2}~ \ln\frac{Q^2}{k_{\bot}^2}
 F(\frac{Q^2}{k_{\bot}^2}).
\eeq %
where $\mu$ is an infrared cut-off.

Differentiates Eq. (\ref{eq:LLA}) with $\mu$, we obtain %
\beq %
\frac{d}{d\ln\mu}F(\frac{Q^2}{\mu^2})=-\frac{\alpha}{\pi}
  \ln\frac{\mu^2}{Q^2}F(\frac{Q^2}{\mu^2})
\eeq %
The above equation coincides with the RG equation (\ref{eq:rge})
with the LL anomalous dimension but the physical meaning is
different. Our derivation starts from the hard function $C(\mu)$
with hard gluons exchange, the above infrared evolution equation
considers the soft gluons exchange.

In the leading double-logarithmic approximation method, we find a
cascade mechanism for Feynman diagrams, the locality of the cascade
scattering and the related evolution equation. These ideas are
analogous to our intuitive understanding about the renormalization
group method although they are expressed in different languages.

At last, we discuss the dependence of the Sudakov form factor on the
regularization methods. In \cite{Sudakov}, the one-loop correction
to the form factor is $-\frac{\alpha}{2\pi}\ln^2\frac{Q^2}{p^2}$ by
using an off-shell regularization method where $p^2$ is the
off-shellness of quarks. The photon mass regularization gives the
one-loop correction as
$-\frac{\alpha}{4\pi}\ln^2\frac{Q^2}{m_{\gamma}^2}$ where
$m_{\gamma}$ is the fictitious mass of photon \cite{Jackiw}. In our
method of regularization, the one-loop result is
$-\frac{\alpha}{8\pi}\ln^2\frac{Q^2}{\delta^2}$ \footnote{We thank
G.P. Korchemsky for pointing out one relation $\delta=\lambda^2/Q$
between our calculations with the results in his paper
\cite{Korchemsky}.}. Different methods give results by a factor of
$2$ or $4$ difference.

\subsection{The Wilson loop method}

Both the Wilson loop method and the CSS method in the next
subsection utilize the RG technic to sum the double logarithms. The
derivations of the evolution equations are done in a different way.
We denote the approach which uses the cusp dimension explicitly as
the Wilson loop method.

The main ingredient in the Wilson loop method is a a renormalization
group equation for a gauge invariant renormalized soft function
$W_R$: %
\beq \label{eq:cusp}%
\left(\mu\frac{\partial}{\partial \mu} +\beta(g)\frac{\partial}
  {\partial g} +\Gamma_{cusp}(\gamma, g)\right)W_R=0.
\eeq %
The derivation of the above equation uses the renormalization
methods given in \cite{Polyakov, BNS, BGSN, KR}. In \cite{Polyakov},
it is pointed out that the vacuum average of a Wilson loop with a
cusp contains extra UV divergences even after the ordinary field
renormalization.  One advantage of the cusp dimension is that it
involves more geometrical meanings.

The on-shell Sudakov form factor is discussed in \cite{Korchemsky}.
The author uses the methods in \cite{KR3} to give a factorized form
which shares some similarities with our operator language. For
example, the coupling of collinear gluons to the hard part vanishes
in the axial gauge, their effects can be included by a gauge
transformation in a general gauge. The final result is similar to
our collinear gauge invariant quantity $W_c^{\dagger}\xi$. A
renormalization group equation for hard function $F_H$ is derived by
using the Eq. (\ref{eq:cusp}) as %
\beq %
\left(\mu\frac{\partial}{\partial \mu} +\beta(g)\frac{\partial}
  {\partial g}\right)\frac{dF_H(Q^2/\mu^2)}{d\ln Q^2}=
  \Gamma_{cusp}(g)F_H(Q^2/\mu^2).
\eeq %
It is consistent with Eq. (\ref{eq:rge}).

In \cite{Korchemsky2}, the off-shell Sudakov form factor is
discussed. There are three scales for the off-shell case: $Q$, $M$
and $M^2/Q$ where $M^2=-p^2$. Our discussion for the on-shell case
needs to be modified and one RG equation is insufficient. For the
off-shell Sudakov form factor in SCET, we need two-step matching:
the first step integrates out the hard mode with momentum
fluctuations of $Q$, the next step integrates out the collinear mode
with momentum fluctuations of $M$. The details about the two-step
matching was discussed in radiative B decays \cite{BHLN}.

\subsection{The CSS method}

We denote the approach given in \cite{CS, CSS2, BottsS, Collins} as
the CSS method. This method is based on the factorization theorem of
pQCD. The intuitive picture behind the factorization is a reduced
graph. The highly off-shell lines with four momenta of order of $Q$
are contracted to points. The reduced graph is constructed by pinch
singular points and it represents a classical scattering process. In
leading power of $1/Q$, the reduced graphs for the quark form factor
contain collinear, soft and hard graphs. In SCET, the highly
off-shell contributions are denoted as the heavy mode and they need
to be integrated out to obtain a low energy effective theory.
Because the two methods describe the same physics, the low energy
physics in the SCET should be exactly equal to the contributions in
the reduced graphs. The method of momentum regions \cite{BS} extends
the reduced graph analysis beyond leading power. The reduced graph
analysis and the method of regions can be used to check that the
SCET reproduces the IR physics of QCD. About the comparison of the
factorizaiton between the CSS method and SCET, some discussions are
given in \cite{B3}.

After separating the quark form factor into collinear, soft and hard
parts, the CSS method differentiates the form factor $F$ with
respect to $\ln Q$ and obtain functions $K$ and $G$ which do not
contain
double-logarithms%
\beq %
\frac{\partial \ln F}{\partial \ln Q}=K(m/\mu,g)+G(Q/\mu,g).
\eeq %
The functions $K, G$ are derived from a gauge dependence of the jet
functions. The RG equations for the CSS method are%
\beq %
\mu\frac{d G}{d\mu}=-\mu\frac{d K}{d\mu}=\gamma_K.
\eeq %
For massless case in QCD, the $K$, $G$ and $\gamma_K$ are
\cite{Collins}%
\beq %
K&=&\frac{\as}{\pi}C_F\frac{1}{\epsilon}; ~~~~~~~~~
G=-\frac{\as}{\pi}C_F\left[\ln\frac{Q^2}{\mu^2}-\frac{3}{2}
  \right]; \non \\
\gamma_K&=& 2\frac{\as}{\pi}C_F +\left[
 \left(\frac{67}{18}-\frac{\pi^2}{6}\right)C_A-
 \frac{10}{9}n_f T_F \right]\left(\frac{\as}{\pi}\right)^2 C_F.
\eeq %

Compared with the results of the last section in SCET, it is
easy to obtain the relations below%
\beq %
G=-\gamma_C, ~~~~~~~~~~~~~~~~\Gamma=\gamma_K~\ln\frac{Q}{\mu}.
\eeq %
The interpretation of the function $G$ as the anomalous dimension
$\gamma_C$ is not accidental because the $G$ function in the CSS
method represents the short-distance contribution.

The position space representation of the CSS method had been applied
into the inclusive processes in \cite{CS, CSS2}, exclusive processes
in \cite{BottsS, LS} and recently into the exclusive B meson decays
in \cite{pQCD}. Because the energy $Q$ is not large enough to ensure
the condition $Q\gg 1/b\gg \lqcd$, the consistency of applying the
perturbative Sudakov form factor is problematic at the experimental
accessible energy regions. This question was addressed in
\cite{DSWY}.

A worldline approach to the Sudakov form factor starts a view closer
to the string theory is discussed in \cite{GKKS}. This approach is
applied into the pion form factor in \cite{SSK}.

\section{Discussions and conclusions}

In this study, we have studied in detail the Sudakov form factor in
the framework of soft-collinear effective field theory.  In the
effective theory, the Sudakov form factor is the coefficient
function running from high to low scale. The exponentiation of the
Sudakov form factor is due to that it is the solution of a
renormalization group equation. To this extent, the renormalization
group equation for the Sudakov form factor is similar to the
evolution equations for the parton distribution functions in deep
inelastic scattering and the evolution equation for the hadron
distribution amplitude. The positive leading-logarithmic anomalous
dimension lead to the suppression of the Sudakov form factor at
large $Q$. We discuss an intuitive picture of the cascade mechanism
behind the renormalization group method.

We compared our method with other approaches for the Sudakov
resummation. The ladder diagrams in the double-logarithmic
approximation method provides an analogy with the intuitive
understanding of the renormalization group method. The Wilson loop
method uses a cusp anomalous dimension which has a clear geometrical
origin. The CSS method gives a factorization of the Sudakov form
factor from diagrammatic analysis and uses two functions to define
the renormalization group evolution. All the methods give the
consistent results. This may indicate that our physical world can be
interpreted from different and complementary points of view. As a
personal opinion, we think that the method of integrating out energy
scales step by step from the effective field theory is more natural
and simpler for the multi-scales problem.

\vspace{0.3cm} $Note ~added:$ After the finish of the paper, we are
informed that a factorization proof of the Sudakov form factor in
SCET was discussed in \cite{B4}. The main difference is that they
use the hybrid position-momentum representation while we will use
the position space formulation.

\section*{Acknowledgments}

It is a pleasure to thank J. Bernabeu, N. Kochelev for useful
discussions and J.C. Collins, G.P. Korchemsky for comments on the
manuscript. The author acknowledges a postdoc fellowship of the
Spanish Ministry of Education. This research has been supported by
Grant FPA/2002-0612 of the Ministry of Science and Technology.

\appendix

\section{}
In this appendix, we provide a detailed calculation of the
one-loop corrections to the quark form factor. We use the
regularization method proposed in \cite{BDS}.

The one-loop vertex correction in the full theory is%
\beq %
\label{eq:fulli}%
I_{full}&=&-ig_s^2 C_F\mu^{\prime 2\epsilon}\int\frac{d^d k}
  {(2\pi)^d} \frac {\gamma_{\rho}(\kslash+\pslash_B)
  \gamma_{\mu}(\kslash+\pslash_A) \gamma^{\rho}}
  {[(k+p_A)^2+i\epsilon][(k+p_B)^2+i\epsilon]
  [k^2-\delta(n_++n_-)\cdot k+i\epsilon]} \non \\
 &=&\frac{\as}{4\pi}C_F \gamma_{\mu} \int_0^1 dx \int_0^{1-x} dy
   \left\{ \Gamma(\epsilon)\left(\frac{4\pi \mu^{\prime 2}}
   {\Delta}\right)^{\epsilon}
   2(1-\epsilon)^2-\frac{2(1-x)(1-y)Q^2}{\Delta} \right \}\non \\
 &=&\frac{\as}{4\pi}C_F \gamma_{\mu}\left[ \frac{1}{\epsilon}
 +{\rm ln}\frac{\mu^2}{Q^2}-\frac{1}{2}{\rm ln}^2\frac{Q^2}{\delta^2}
 +2{\rm ln}\frac{Q^2}{\delta^2}-\frac{\pi^2}{3} \right]
\eeq %
where $\Delta=xyQ^2-(x+y)(1-x-y)Q\delta+(1-x-y)^2\delta^2$.

The collinear-to-A contribution is%
\beq \label{eq:JA}%
J_A&=&-ig_s^2 C_F\mu^{\prime 2\epsilon}\int\frac{d^d k}
   {(2\pi)^d} \frac {-\gamma_{\mu}2(p_A^+-k^+)}
   {[k^2-2p_A\cdot k+i\epsilon]~(n_-\cdot k)~
   [k^2-\delta (n_-\cdot k)+i\epsilon]} \non \\
 &=&-\frac{g_s^2}{2\pi}C_F \mu^{\prime 2\epsilon}\gamma_{\mu}
   \int_0^Q \frac{d k^+}{k^+}(p_A^+-k^+)\int \frac{d^{d-2} k_{\bot}}
   {(2\pi)^{d-2}}\frac{1}{k_{\bot}^2+\delta k^+(1-\frac{k^+}{p_A^+})}
  \non \\
 &=&\frac{\as}{4\pi}C_F \gamma_{\mu}\left[ \frac{2}{\epsilon^2}
   +\frac{2+2{\rm ln}\frac{\mu^2}{Q \delta}}{\epsilon}
   +{\rm ln}^2\frac{\mu^2}{Q\delta}+2{\rm ln}\frac{\mu^2}{Q\delta}
   +4-\frac{\pi^2}{6} \right]
\eeq %
From the first to the second line of the above equation, we
perform the $k^-$ integral first by closing the contour in the
lower half plane. The $k^-$ pole is chosen as
$k^-=\frac{k_{\bot}^2+\delta k^+-i\epsilon}{k^+}$ and the range
$k^+$ is $0<k^+<p_A^+$. The divergence of $k_{\bot}\to \infty$ is
regulated by choosing dimension $d<4$. There is another
singularity coming from the momentum region $k^+\to 0$. The
overlapping of the two singularities lead to the double poles
$\frac{1}{\epsilon^2}$.  The result of the collinear-to-B
contribution $J_B$ is the same as $J_A$.

The soft contribution is %
\beq \label{eq:S}%
S&=&-ig_s^2 C_F\mu^{\prime 2\epsilon}\int\frac{d^d k}{(2\pi)^d}
  \frac{ 2\gamma_{\mu} }{[n_-\cdot k+i\epsilon][n_+\cdot k+i\epsilon]
  [k^2-\delta (n_++n_-)\cdot k+i\epsilon]} \non \\
  &=& -\frac{g_s^2}{2\pi}C_F \mu^{\prime 2\epsilon}\gamma_{\mu}
   \int_{\delta}^{\infty}\frac{dk^+}{k^+}\int \frac{d^{d-2} k_{\bot}}
   {(2\pi)^{d-2}}\frac{1}{k_{\bot}^2+\delta k^+} \non \\
  &=&\frac{\as}{4\pi}C_F \gamma_{\mu}\left[ -\frac{2}{\epsilon^2}
  -\frac{2{\rm ln}\frac{\mu^2}{\delta^2}}{\epsilon}
  -{\rm ln}^2\frac{\mu^2}{\delta^2}-\frac{\pi^2}{6} \right]
\eeq %
The contour of the $k^-$ integral is closed in the lower half
plane and choose the pole at $k^-=\frac{k_{\bot}^2+\delta
k^+-i\epsilon}{k^+-\delta}$ with $k^+>\delta$. The soft function
is Q-independent.


\end{document}